\documentclass[a4paper,USenglish,cleveref, autoref, thm-restate]{lipics-v2021}

\usepackage{graphicx}
\usepackage{amsmath}
\usepackage{amssymb}
\usepackage{ifthen}
\usepackage{caption}
\usepackage{subcaption}
\usepackage{hyperref}

\usepackage[dvipsnames]{xcolor}
\usepackage{tikz}
\usetikzlibrary{arrows,%
                calc,%
                snakes,positioning,chains,
                shapes.geometric}%
\usetikzlibrary{decorations.pathmorphing}

\newcommand{\Nat}{\mathbb{N}}

\def\idtt#1{\ensuremath{\mathtt{#1}}}

\newcommand{\np}{\ensuremath{\mathrm{NP}}}

\newcommand{\paraTC}{\ensuremath{\mathrm{para\!\!-\!\!TC}}}

\newcommand{\paraAC}{\ensuremath{\mathrm{para\!\!-\!\!AC}}}
\newcommand{\paraL}{\ensuremath{\mathrm{para\!\!-\!\!L}}}
\newcommand{\xl}{\ensuremath{\mathrm{XL}}}

\def\op#1{\idtt{#1}}
\def\deg{\idtt{deg}}
\DeclareMathOperator{\poly}{poly}

\definecolor{lightblue}{rgb}{0.1,0.1,1}

\usepackage{todonotes}

\usepackage{amsfonts}


\nolinenumbers 

\hideLIPIcs

\title{Space-Efficient Graph Kernelizations}

\titlerunning{Space-Efficient Graph Kernelizations}

\author{Frank Kammer}{THM, University of Applied Sciences Mittelhessen, Giessen, Germany}{Frank.Kammer@mni.thm.de}{https://orcid.org/0000-0002-2662-3471}{}

\author{Andrej Sajenko}{THM, University of Applied Sciences Mittelhessen, Giessen, Germany}{Andrej.Sajenko@mni.thm.de}{https://orcid.org/0000-0001-5946-8087}{DFG -- 379157101.}

\authorrunning{F. Kammer and A. Sajenko}
\Copyright{Frank Kammer and Andrej Sajenko}


\relatedversion{This is the full version of the paper~\cite{KamS24}, which includes  proofs and an additional result to cluster editing and deletion.}

\begin{document}

\maketitle  

\begin{abstract}
Let $n$ be the size of a parameterized problem and $k$ the parameter. 
We present kernels for \textsc{Feedback Vertex Set}, \textsc{Path Contraction}
and \textsc{Cluster Editing/Deletion} whose sizes are all polynomial 
in $k$ and that are computable in polynomial time and 
with $O(\poly(k) \log n)$ bits (of working memory). 
By using kernel cascades, we obtain the best known 
kernels in polynomial time with $O(\poly(k) \log n)$ bits.

\end{abstract}

\section{Introduction}
\setcounter{page}{1}
With the rise of big data the focus on algorithms
that treat space as a valuable resource becomes increasingly important.
Large inputs may cause ``standard'' solutions to fail their execution due to out-of-memory errors, or cause them to spend a significant amount of time for memory swapping due to cache faults.

Within the last ten years, there is a new research direction 
called {\em space-efficient algorithms}
where one tries to solve a problem with as little space as possible while 
``almost'' maintaining the same running
time of a standard solution for the problem 
under consideration.
Space-efficient algorithms are 
mostly designed for problems
that already run in polynomial time,i.e., we have algorithms for
connectivity problems~\cite{ChakSS20,ElmHK15,Hag20},
matching~\cite{DatKM16} and 
other graph problems~\cite{HagKL18,KamKL16}.
Further algorithms are known for, e.g., 
sorting~\cite{Bea91,PagR98} and geometric
problems~\cite{AsaBBKMRS14,BarKLSS13}.

Our goal is to combine the research on space-efficient algorithms 
with parameterized algorithms.
In the classical literature a 
{\em parameterized problem} $P \subseteq \Sigma^* \times \Nat$ is a language
where $\Sigma$ is a finite alphabet and the second part $k\in \Nat$ is
called {\em parameter}.
In addition, $P$ is called {\em fixed-parameter tractable} (FPT)
if there exists an algorithm $\mathcal A$ (called {\em FPT algorithm})
and a computable function $f$
such that, given an {\em instance} $(I, k) \in \Sigma^* \times \Nat$,
the algorithm $\mathcal A$ correctly decides whether $(I, k) \in P$
in a time bounded by $f(k) \poly(|I|)$.
A popular way to find an FPT algorithm is to find a so-called kernelization algorithm.
Given a parameterized problem $P$, a {\em kernelization algorithm}
(or simply called {\em kernelization})
is a polynomial-time algorithm $\mathcal A: (\Sigma^* \times \Nat) \rightarrow (\Sigma^* \times \Nat)$ such that, 
given an instance $(I, k) \in \Sigma^* \times \Nat$, then $(I', k') = \mathcal A(I, k)$
is another instance (called {\em kernel}) of the problem 
with the property that $(I', k') \in P$ if and only if $(I, k) \in P$ and $|I'|, k' \le g(k)$ (where usually $k' \le k$) 
for some function~$g$.
Then the kernel can usually be solved by a brute-force algorithm in $O(f(g(k))$ time, for some function~$f$.
Furthermore, we call $(I, k)$ a {\em yes}-instance exactly if $(I, k) \in P$. 
Otherwise, we call it a {\em no}-instance.
A yes- or no-instance of constant size is called {\em trivial}.
The modification steps of the kernelization are called {\em reduction rules} and
such a rule is {\em safe} if, given input $(I, k)$, it produces an output $(I', k')$ such that 
	$(I', k')$ is a yes-instance if and only if $(I, k)$ is a yes-instance.

In this paper, we describe {\em space-efficient kernelizations}, 
which we define as a kernelization $\cal A$ (as above)
with the additional property of using $O(h(k) \log |I|)$ bits for some computable function~$h$.
Following this definition, a {\em space-efficient FPT algorithm}
is an FPT algorithm that runs within $O(h(k) \log |I|)$ bits of space.
By using our kernel of $g(k)$ vertices and edges as an intermediate kernel, which needs 
$O(g(k) \log |I|)$ bits to be stored, we then can apply the best known 
solution to further reduce the kernel size via {\em kernel cascades} 
(i.e., the consecutive application of a kernelization on the kernel).
Alternatively, we can use the kernel to easily build a space-efficient FPT algorithm 
$\mathcal A$ that produces an optimal solution $S$ for the instance $(I, k)$
in time $f(k) \poly(|I|)$ using $O((g(k) + h(k)) \log |I|)$ bits for some functions $f$ and $h$.
We focus on graph problems, i.e.,
an instance $I$ is a graph $G = (V, E)$ as well as $|I| = |V| + |E|$, and
our space-efficient kernelizations are {\em space-efficient graph kernelizations}.
For the remainder of the paper, let $n$ and $m$ be the \mbox{vertices and
edges, respectively, of the graph under consideration.}
Note that for simple graphs, $O(\log n) = O(\log m)$ holds.

We also recognize the need for {\em full kernels}~\cite{Dam06}, i.e., 
a kernel that contains the vertices/edges of all minimal solutions in 
a yes-instance $(G, k)$.
Such a full kernel
allows us to enumerate all minimal solutions of size at most~$k$.
Those kernels are, e.g., necessary for the application of frameworks
such as shown in~\cite{HegHKNRA21} and for parameterized enumeration~\cite{Dam06}.
Our computation model is based on a read-only word-RAM with a 
word size of $w = \Omega(\log n)$ bits, enabling constant-time arithmetic 
operations ($+, -, \cdot, /$, modulo) and bit-shift operations on $w$-bit sequences. 
The input is divided into three types: read-only {\em input memory}, write-only 
{\em output memory}, 
and read-write {\em working memory}. Space-efficient algorithm space bounds are 
typically in bits and refer to the working memory. When expressed in words, 
these space bounds include an extra factor of $\Theta(\log w)$ or $\Theta(\log n)$ 
depending on the specific implementation, making them less precise or 
more complex to describe. In contrast, we express kernel sizes in 
terms of vertices/edges (i.e., in words), following the conventional 
approach for describing kernel size.

{\bfseries Parameterized Space Complexity.} 
In classical research, the focus is often on achieving minimal space bounds, 
but this comes at the cost of significantly increased running times, 
rendering them impractical.
{\em Parameterized space complexity}
is a research area where one mainly
classifies problems based on the amount of memory
required to solve them.  Two important classes in this field are $\paraL$
(aka.  logspace + advice) and $\xl$ (aka.  slicewise logspace).  $\paraL$
contains problems that can be solved with $f(k) + O(\log n)$ bits, while
$\xl$ contains problems that can be solved with 
$O(f(k) \log n)$
bits~\cite{CaiCDF97}.

We next give an overview over parameterized space complexity restricted to graph problems.
An early work on parameterized space complexity is due to
Cai et al.~\cite{CaiCDF97} who showed that vertex cover is in 
$\paraL$.
Flum and Grohe~\cite{FlumG02} presented that model-checking 
problems of first-order formulas of bounded degree graphs 
are also in $\paraL$.
Elberfeld et al.~\cite{ElbST15} showed that \textsc{Feedback Vertex Set} (FVS) is in $\xl$.
Bannach et al.~\cite{BanST15} studied packing, covering
and clustering problems and show (among other results)
that \textsc{Triangle Packing}, \textsc{(Exact Partial) Vertex Cover},
and \textsc{Many Cluster Editing} are in $\paraL$ (more precisely in
a class called $\paraAC^0 \subseteq \paraL$)
and \textsc{Cluster Editing} is in $\paraL$ (more precisely in $\paraTC^0 \subseteq \paraL$).
Fafianie and Kratsch~\cite{FafK15} showed that several graph deletion 
problems where the target classes have finite forbidden 
sets are also in $\paraL$. 
This result was recently 
generalized to infinite forbidden sets by Biswas et al.~\cite{BisRS21}, 
who showed that deletion problems like deletion to linear 
forest and deletion to pathwidth~1 are also in $\paraL$.

To our knowledge neither the membership of $\paraL$ nor a 
lower bound for FVS was discovered yet.
While FVS can be formulated as a deletion problem, Biswas et al.~\cite{BisRS21} 
mentioned that the techniques required are not ``easily'' 
applicable to FVS (and other deletion problems). 
They attacked the problem by using different parameterizations 
and showed that (among other problems) FVS
is in $\paraL$ if parameterized by vertex cover.
Moreover, they presented a space-efficient FVS algorithm, 
parameterized by solution size $k$, that runs in $5^k n^{O(1)}$
time and with $O(k \log n)$ bits, an improvement over the 
previously known best space-efficient FVS algorithm of Elberfeld 
et al.~\cite{ElbST15}, which runs in $O(k^kn^5)$ time 
and uses $O(k \log n)$ bits.

{\bfseries Approach and Contribution.}
Our approach departs from the conventional method of computing a 
kernel by applying reduction rules globally to the entire graph, 
a process that can be resource-intensive in terms of space. 
Aiming for a memory usage of $O(\poly(k) \log n)$ bits, 
we show a process that involves systematically modifying and condensing 
disjoint subgraphs of the input graph $G = (V, E)$ while preserving $\poly(k)$ 
vertices and edges in the resulting kernel.
The main idea here is to make use of a separator set $U$ of size $\poly(k)$, 
which partitions graph $G$ into disjoint subgraphs. 
Each subgraph undergoes a reduction process under consideration
of the separator $U$ to efficiently shrink it to $\poly(k)$ vertices and edges.
Subsequently the reduced subgraph is then carefully 
integrated into an initial kernel $G' = G[U]$ under construction.
$G'$ is then repeatedly reduced to ensure we stay within our space bound.
To make our algorithm work we need to show that both, computing the separator and applying the reduction rules, must be implemented
with $O(\poly(k) \log n)$ bits. This often means that we are not able to
\mbox{run all known reduction rules or to run the rules in a restricted
setting.}

In Section~\ref{sec:pc}, we present a simple \textsc{Path-Contraction} kernelization
that runs in
 $O(n \log k + \poly(k))$ time using $O(\poly(k) \log n)$ bits.
To find a kernel for \textsc{Path Contraction}, 
one usually searches for {\em bridges} (i.e., edges whose removal disconnect the graph) 
and merges the 
endpoints of such a bridge to a single vertex~\cite{HegHLLP14,LiFCH17}.
Bridges are usually found by running a DFS---to the authors knowledge, 
all polynomial-time DFS
need $\Omega(\sqrt{n})$ bits and a polynomial-time depth-first search (DFS) 
with $O(\sqrt{n})$ bits is due to Izumi and Otachi~\cite{IzuO20}. 
Instead of this reduction rule,
we use a separator $U$ (which is the queue of a breadth-first search (BFS))
and iteratively expand our kernel while shrinking 
{\em induced degree-2-chains} (paths whose vertices all have degree~2) as long as they consist of more than $k+1$ edges.
To achieve our space bound we 
show that a yes-instance of \textsc{Path Contraction} cannot have
a tree as an induced subgraph with more than $k + 2$ leaves.
This bounds the size of $U$ by $O(k)$ and makes it possible to construct a BFS algorithm that
stores at most $O(k)$ vertices at a time,
allowing us to construct a kernel with $O(k^2)$ vertices and edges
in $O((n + k^2) \log k)$ time using $O(k^2 \log n)$ bits.
To get the current best kernel size we subsequently apply Li et al.'s~\cite{LiFCH17} 
polynomial-time kernelization and so get a kernel of $3k + 4$ vertices
in $O(n \log k + \poly(k))$ time and using $O(\poly(k) \log n)$ bits.
Li et al.'s kernelization for \textsc{Path Contraction} builds on Hegernes et al.'s kernelization~\cite{HegHLLP14} and uses $\Theta((n + m) \log n)$ bits due to searches for bridges
and to store the modification 
in adherence to their reduction rules.

Our main result is a new kernelization for \textsc{Feedback Vertex Set} (FVS) 
in Section~\ref{sec:fvs}, prefaced by its own preliminaries in  Section~\ref{sec:treeTraversalSection}.
Our idea is to compute
an approximate minimum feedback vertex set $U$ as a separator
 whose removal partitions the graph into several trees. 
 We use a so-called Loop Rule and a restricted Flower Rule 
as well as a so-called Leaf Rule and Chain Rule.
If the kernel is still too
large, we follow ideas from Thomass\'e~\cite{Tho10}. 
For details on these rules, see Section~\ref{sec:fvs}.
We want to remark that a 
solution for \textsc{Path Contraction} of size $k$ implies one for \textsc{FVS}
of size $2k$ by simply taking the endpoints of the contracted edges.
Our kernelization for FVS 
runs in $O(n^5 \poly(k))$ time, uses $O(k^{4} \log n)$ bits
and outputs a kernel of $n' = 2k^2 + k$ vertices. 
Note 
that \textsc{Feedback Vertex Set} has no kernel of size $O(k^{2-\epsilon})$
for any constant $\epsilon > 0$ unless $\np \subseteq \mathrm{coNP/poly}$~\cite{DellM14}.
After computing our kernel we can use the deterministic algorithm of
Iwata and Kobayashi to solve it in $O(3.46^k n') = O(3.5^{k})$
time or the randomized algorithm of Li and Nederlof
to solve it in $2.7^k \mathrm{poly}(n')$ time.
In total, we can solve FVS in $O(n^5 \poly(k) + 3.5^{k})$ time and with $O(k^{4} \log n)$ bits. 

Li and Nederlof and Iwata and Kobayashi
do not focus on space efficiency 
in their search-tree algorithms for \textsc{Feedback Vertex Set}
and thus do not state a space bound.
Based in their
description they assume either to be able to
modify the input graph or to create at least one copy of it,
which gives us a lower bound of $\Omega((n + m) \log n)$ bits.
By assuming that one stores
a copy of the reduced graph whenever descending in the search tree,
the two search-tree algorithms have a space
upper bound of $O(k(n + m) \log n)$ bits.

Compared to Elberfeld et al.'s algorithm~\cite{ElbST15}, which solves FVS in time $O(k^kn^5)$, we are faster, but we use $O(k^{4} \log n)$ instead of $O(k \log n)$ bits.
Concurrently and independently to our result, Biswas et al.~\cite{BisRS21} presented an iterative compression algorithm (based on the Chen et al.~\cite{ChenFLLV08} algorithm) that maintains the space bound of $O(k \log n)$ bits and has a runtime of $O(5^k n^{O(1)})$.
The degree of the polynomial is not mentioned explicitly, 
presumably due to the fact that some of the used log-space auxiliary results do not mention their ``exact'' running time either.
However, we and they use the $O(\log n)$ bit realizations of the so-called Leaf and Chain Rules (aka.\ Degree-2 Rule) from Elberfeld et al.~\cite{ElbST15}.
The non space-efficient version of the rules removes vertices from the input graph that are not relevant for solving FVS, but due to the space restriction,
the information of the graph resulting from the removal is computed on demand. By our analysis (Lemma~\ref{lem:spaceBasicRule1and2}) this results in a running time factor of $\Theta(n^5 k \log k)$ for each vertex / edge access.
Moreover, Biswas et al. have a nested loop where $\Theta(n^2)$ connectivity tests have to be performed, which increases the running time to
$\Theta(n^7 \poly(k) + 5^k n^{\Theta(1)})$.
Based on that our running time is faster in $k$ and in $n$, but they use only $O(k \log n)$ bits. A summary is shown in Table~\ref{tab:fvs-comparison}.

\begin{table}[t]
\centering
\begin{tabular}{r|c|c}
  Authors & Time & Space [bits]\\
  \hline
  (randomized) Li and Nederlof~\cite{LiN20} & $O(2.7^k(n + m))$ & $\Omega((n + m) \log n)$ \\
  Iwata and Kobayashi~\cite{IwataK19} & $O(3.46^kn)$ & $\Omega((n 
 + m) \log n)$ \\
  Elberfeld et al.~\cite{ElbST15} & $O(k^kn^5)$ & $O(k \log n)$ \\
  Biswas et al.\ \cite{BisRS21} & $O(n^7 \poly(k) + 5^k n^{\Theta(1)})$ & $O(k \log n)$ \\
  This paper + Iwata and Kobayashi~\cite{IwataK19} & $O(n^5 \poly(k) + 3.5^{k})$ & $O(k^{4} \log n)$\\
  \hline
\end{tabular}
\vspace{1mm}
\caption{A time and space comparison of \textsc{Feedback Vertex Set} algorithms.}\label{tab:fvs-comparison}
\end{table}

In Section~\ref{sec:ce}, 
we adapt a standard technique
for \textsc{Cluster Editing/Deletion} to compute a full kernel
of $O(k^2)$ vertices in $O(nm \log k)$ time and within $O(k^2 \log n)$ bits.
Here we use $O(k^2)$ times a rule searching for so-called
{\em conflict triples}.
This easily allows us to bound the size of a separator $U$ by $O(k^2)$.

\section{Path Contraction}\label{sec:pc}
Let $G$ be an $n$-node $m$-edge graph
and let $C$ be a subset of edges of $G$.
We write $G/C$ for the graph obtained from $G$ by contracting each edge in $C$.
{\em Contracting an edge} is done by merging its endpoints 
and removing any loops or parallel edges afterwards.
In the parameterized \textsc{Path Contraction} problem, a connected graph $G = (V, E)$ is given
together with a parameter $k$ and the task
is to find a set $C \subseteq E$ with $|C| \le k$ such that $G/C$ is a path. 
In particular, $G/C$ is a connected graph with $n' \in \Nat$ vertices and $n'-1$ edges.
One reduction rule used in Li et al.'s~\cite{LiFCH17} and
in Heggernes et al.'s~\cite{HegHLLP14} kernelization is an iterative contraction of a
bridge for which no polynomial-time $O(\poly(k) \log n)$-bits
algorithm is known.
(Bridges are found by running a DFS and 
the best-known polynomial-time DFS with a minimum of space 
uses $\Omega(\sqrt n)$ bits~\cite{IzuO20}.)
Instead of computing bridges, we introduce 
two new reduction rules below.
In the following, a {\em subtree} $T$ of $G$ is a subgraph of $G$ that is a tree.
Moreover, let a {\em degree-2 chain} be a maximal simple path
$P = v_1, \ldots, v_{\ell}$ 
($\ell \in \Nat$) whose 
vertices $v_1, \ldots, v_{\ell}$ are all of degree~$2$.
Observe, if $P$ is not a cycle,
then $v_1$ and $v_\ell$ must each have a neighbor that is not of degree~2.

\begin{description}
\item[Rule 1] If there exists a degree-2 chain $P$ with more than $k + 1$ edges, 
contract all except $k + 1$ arbitrary edges of $P$.
  
\item[Rule 2]
After $k$ applications of Rule 1, 
if $G$ contains more than 
$k^2 + 4k + 1$ vertices, 
more than $(3k^2 + 13k)/2$
edges, or a subtree with more than 
$k+2$ leaves, then output ``no-instance''.
\end{description}

The bound on the number of leaves in Rule 2
helps us to guarantee that our kernelization
works in our space bound.
We want to remark that we do not apply Rule 2 exhaustively; 
more precisely, we do not explicitly search for subtrees 
with more than~$k+2$ leaves, which is NP-hard.

\begin{lemma}\label{lem:pathContractionRule2Safe}
Rule 1 and 2 are safe and produce a full kernel.
\end{lemma}
\begin{proof}
\textbf{Rule 1 is safe.}
Observe that a solution either contracts the entire degree-2 chain $P$ or none of it. 
Since contracting an entire degree-2 chain with more than $k+1$ edges is not allowed, 
any solution does not contract any edge of $P$. 
Thus, Rule 1 is safe and we do not remove an edge of any minimal solution.

\textbf{Rule 2 is safe. Vertex bound.}
Assume that $(G, k)$ is a yes-instance. 
Let us define~$C$ 
as the subset of edges from $E(G)$ 
that are requisite for contraction to 
arrive at a solution for $G$.
Since $C$ is a solution for $G$
it is also a valid solution for $G'$.
Consider the path $P^*$, 
defined by $P^* = G' / C$.
Observe that $P^*$ is an 
amalgamation of subpaths, each limited 
to $k + 1$ edges. 
These subpaths, when traced back to 
$G'$, are individually connected by 
a unique vertex with a 
minimum degree of~3.
Contraction on this specific vertex is 
obligatory to obtain the path $P^*$.
Since one edge contraction
reduces the vertex count by one, 
the path $P^*$ would comprise a vertex 
count less than that of $G'$ by at 
most $k$ vertices. 
Consequently, there is an upper 
limit of 
$k$ for such unique vertices in $P^*$. 
This means that $P^*$ can accommodate at 
most $k + 1$ of these subpaths.

From the above deductions, it can be derived that the vertex count for
$P^*$ is capped at $(k + 1)^2 + k$. Extending this reasoning, 
$G'$ has a maximum vertex count of $(k + 1)^2 + 2k$.

\textbf{Edge bound.}
Let $N(u)$ be the neighbors of a vertex $u$.
Note that by definition, a contraction of an edge $\{u, v\}$ reduces 
the number of vertices by one, and reduces the number of 
edges by $|N(u) \cap N(v)| + 1$.
Since $G / C$ is a path where each vertex is of maximal degree~2, 
$G$ cannot have a vertex $u$ with $|N(u)| \ge k + 3$ 
since at least $k + 1$ contractions are 
needed to remove $k + 1$ neighbors of $u$ to reach degree at most~2.
Hence, the common neighborhood of two adjecent vertices is at most $k + 1$.
Therefore, if we have $k' \le k$ contractions left, we can remove
at most $k' + 2$ edges with one contraction.
Thus, with $k$ possible contraction we can remove at most
$\sum_{i=1}^k (i + 2) = 2k + \sum_{i=1}^k i = \frac{1}{2}(k^2 + k) + 2k$ edges.
Hence, $G$ cannot have more than $n' - 1 + \frac{1}{2}(k^2 + 5k)$
edges in total.
Summarized, the kernel consists of $O(k^2)$ vertices and edges and is a full kernel
since the only modification are done by Rule~1.

\textbf{Bound on leaves in subtrees.}
Let us say that a {\em super vertex} $w$ is a vertex obtained from a contraction of
({\em normal}) vertices $u$ and
$v$ of $G$. For an easy intuition, we then say that $w$ {\em contains} $u$ and $v$.
Consider the following fact: if $G'$ is obtained from $G$ by contracting an arbitrary number of
edges and $(G, k)$ is a yes-instance, then $(G',k)$ is also a yes-instance. 
By induction, it
suffices to show the fact for graphs $G'$ that are obtained by one
contraction of an arbitrary edge~$\{u, v\}$.
Assume that $G' = G/\{\{u,v\}\}$.
Consider an optimal solution $C$ for $(G, k)$ and the path $P = G/C$.
We denote by $S(u)$ and $S(v)$ the super vertices in $G/C$ containing
$u$ and $v$, respectively.
Note that,
either $S(u) = S(v)$ or $S(u)$ and $S(v)$ are
adjacent in $P$.
If $S(u) \ne S(v)$,
then $G'/C = (G/\{\{u,v\}\})/C = (G/C)/\{\{S(u),S(v)\}\} = P/\{\{S(u),S(v)\}\}$ is a path and thus, $C$ is a solution for $(G', k)$.
If $S(u) = S(v)$, then there is a $u$-$v$-path $P'$ in $G'$ consisting only
of edges in $C$.
Let $e$ be an edge of $P'$ and $C' = C \setminus \{e\}$.
Then $G'/C' = G/C$ is a path and $C'$ is a solution $L$ 
for $(G', k - 1)$ and thus for $(G', k)$.
To sum up, the fact holds.

Assume for a contradiction that $G$ has a subtree $T$ with $\ell>k+2$
leaves.
By contracting all edges of $T$ in $G$ without edges that have an endpoint in $L$
we obtain a graph $G'$ and by the fact above $(G', k)$ is a yes-instance.
However, $G'$ has a subtree with $\ell > k + 2$ leaves---a contradiction being a yes-instance. 
Since we only reject no-instances the kernel remains a full kernel. 
\end{proof}

For the kernel construction we use a BFS.
We shortly sketch a usual BFS and the construction of a so-called {\em BFS tree}.
The BFS visits the vertices of an input graph round-wise. 
As a preparation of the first round it puts some vertex $v$
into a queue $Q$, marks it as visited, and starts a round.
In a round it dequeues every vertex $u$ of $Q$, 
and marks $u$ as visited.
Moreover, it puts every unvisited neighbor $w \in N(u)$ of $u$
into a queue $Q'$ and marks it as visited.
We then say that $w$ was first discovered from $u$ and 
add the edge $\{u, w\}$ to an initial empty BFS tree.
If $Q'$ is empty at the end of the round, the BFS finishes. 
Otherwise, it proceeds with the next round with $Q := Q'$
and $Q' := \emptyset$.

During each BFS iteration, the BFS queue $Q$ inherently 
acts as a separator, dividing the graph into two distinct categories:
vertices already encountered by the BFS and those yet to be encountered.
To understand this, envision a BFS queue $Q$ established after a 
given round, prior to the initiation of the subsequent round. 
Let $V_1$ denote the vertices in $G$ already encountered 
by the BFS, excluding those in $Q$, and $V_2$ represent 
vertices yet to be visited. If $Q$ were not a separator, 
an edge $(u, v)$ would exist such that $u$ belongs 
to $V_1$ and $v$ to $V_2$, effectively suggesting 
that the BFS traversal overlooked an unvisited vertex $v$ 
while exploring $u$'s neighbors. This contradicts the BFS algorithm, 
as $v$ should have been incorporated into $Q$.
Therefore, $Q$ functions as the desired separator~$U$.

As the BFS progresses, its exploration adheres to the 
subtree structure of $G$. The BFS queue size is restricted to a 
maximum of $k + 2$ vertices; otherwise, we can determine a 
no-instance by Rule 2 and halt the BFS process. 
With regards to marking vertices as visited, 
since $Q$ effectively delineates between previously visited 
and unvisited vertices after each BFS round, only the last 
BFS round's queue is necessary to verify the visited vertices.
Notably, in this context, separator $U$ consists of vertices in $Q$ and is thus dynamic, 
adapting with each BFS iteration.
So far this approach ensures that every yes-instances is 
traversed and any no-instance is identified in $O(n \log k)$ 
time utilizing only $O(k \log n)$ bits.

\begin{figure}[t]
\centering
\begin{tikzpicture}[scale=0.9,transform shape, every node/.style={on chain,draw,circle, fill=white, inner sep=0pt,
	minimum size=7pt},start chain = going right,node distance = 10pt]

	\foreach \x in {1,...,5} {%
		\node[] (va\x) {};%
	}
	\foreach \i [evaluate=\i as \j using {int(\i+1)}] in {1,...,4} {%
		\draw[] (va\i) -- (va\j);%
	}
	
	\node[yshift=15pt] (vb1) {};%
	\foreach \x in {2,...,3} {%
		\node[] (vb\x) {};%
	}
	\foreach \x in {4,...,6} {%
		\node[dotted] (vb\x) {};%
	}
	\foreach \i [evaluate=\i as \j using {int(\i+1)}] in {1,...,2} {%
		\draw[] (vb\i) -- (vb\j);%
	}
	\foreach \i [evaluate=\i as \j using {int(\i+1)}] in {2,...,5} {%
		\draw[dotted] (vb\i) -- (vb\j);%
	}

	\node[below=20pt of vb1] (vc1) {};%
	\foreach \x in {2,...,3} {%
		\node[] (vc\x) {};%
	}
	
	\foreach \x in {4,5, 6} {%
		\node[dotted] (vc\x) {};%
	}
	\foreach \i [evaluate=\i as \j using {int(\i+1)}] in {1,...,2} {%
		\draw[] (vc\i) -- (vc\j);%
	}
	\foreach \i [evaluate=\i as \j using {int(\i+1)}] in {3,...,5} {%
		\draw[dotted] (vc\i) -- (vc\j);%
	}
	
	\node[right=115pt of va5] (vd1) {};%
	
	\node[yshift=27pt] (ve11) {};%
	\foreach \x in {2,...,5} {%
		\node[] (ve1\x) {};%
	}
	\foreach \x in {6,...,10} {%
		\node[dotted] (ve1\x) {};%
	}
	\foreach \i [evaluate=\i as \j using {int(\i+1)}] in {1,...,4} {%
		\draw[] (ve1\i) -- (ve1\j);%
	}
	\foreach \i [evaluate=\i as \j using {int(\i+1)}] in {4,...,9} {%
		\draw[dotted] (ve1\i) -- (ve1\j);%
	}
	
	\node[below=10pt of ve11] (ve21) {};%
	\foreach \x in {2,...,6} {%
		\node[] (ve2\x) {};%
	}
	\foreach \x in {7,...,10} {%
		\node[dotted] (ve2\x) {};%
	}
	\foreach \i [evaluate=\i as \j using {int(\i+1)}] in {1,...,5} {%
		\draw[] (ve2\i) -- (ve2\j);%
	}
	\foreach \i [evaluate=\i as \j using {int(\i+1)}] in {6,...,9} {%
		\draw[dotted] (ve2\i) -- (ve2\j);%
	}

	\node[below=10pt of ve21] (ve31) {};%
	\foreach \x in {2,...,3} {%
		\node[] (ve3\x) {};%
	}
	\foreach \x in {4,...,5} {%
		\node[dotted] (ve3\x) {};%
	}
	\foreach \x in {6,...,10} {%
		\node[] (ve3\x) {};%
	}
	\foreach \i [evaluate=\i as \j using {int(\i+1)}] in {1,...,9} {%
		\draw[] (ve3\i) -- (ve3\j);%
	}
	
	\node[below=10pt of ve31] (ve41) {};%
	\foreach \x in {2,...,3} {%
		\node[] (ve4\x) {};%
	}
	\foreach \i [evaluate=\i as \j using {int(\i+1)}] in {1,...,2} {%
		\draw[] (ve4\i) -- (ve4\j);%
	}
	
	\draw[] (va5) -- (vb1);%
	\draw[] (va5) -- (vc1);%
	\draw[dotted] (vc6) -- (vd1);%
	\draw[] (vd1) -- (ve11);%
	\draw[] (vd1) -- (ve21);%
	\draw[] (vd1) -- (ve31);%
	\draw[] (vd1) -- (ve41);%
	
	\draw[dashed] (ve13) -- (ve24);%
	
	\draw[dashed] (ve25) -- (ve36);%
	
	\draw[dotted] (ve110) -- (ve210);
	
	\draw[ultra thick] (vc3) to[out=30,in=180] (vd1);%
	\draw[ultra thick] (ve26) to[out=150,in=-30] (ve15);%
	\draw[ultra thick] (ve33) to[out=30,in=150] (ve36);%

	\node also [label={[label distance=3pt]230:\large$v_1$}] (vc1);
	\node also [label={[label distance=3pt]220:$\ldots$}] (vc2);
	\node also [label={[label distance=1pt]220:\large$v^*$}] (vc3);
	\node also [label={[label distance=6pt]220:\large$p$}] (vc6);
	\node also [label={[label distance=-1pt]270:$\ldots$}] (vc4);
	\node also [label={[label distance=6pt]220:\large$v$}] (vd1);
\end{tikzpicture}
\caption{
Our adapted BFS starts from the leftmost vertex, removing dotted 
vertices on a degree-2 chain with over $k + 1$ predecessors 
and connecting the neighbors of removed vertices with bold edges. 
Dashed edges are skipped by the BFS.
}\label{fig:pc}
\end{figure}

To realize Rule 1 we need additional information.
Instead of storing just vertices $v$ on the BFS queue we store quadruples
that we use to identify degree-2 chains and apply Rule~1---see also Fig.~\ref{fig:pc}.
Each quadruple $(v, p, i, v^*)$ consists
of
a vertex $v$ and its predecessor $p$ if $v$ is not the root,
the counter $i \in \{0, \ldots, n\}$ with $i > 0$
	being $v$'s position on a degree-2 chain,
and the vertex $v^*$ with $v^* \ne \op{null}$ being the $(k + 1)$th vertex on a degree-2 chain that contains $v$.
So we can easily check Rule 1 as shown in the proof of Theorem~\ref{th:path-contraction}.
By Rule~2, we can guarantee
our space bound by maintaining $O(k^2)$ vertices and edges.

\begin{theorem}\label{th:path-contraction}
	Given an $n$-vertex instance $(G, k)$ of \textsc{Path Contraction}, there is an
	$O((n + k^2) \log k)$-time
	$O(k^2 \log n)$-bits
	kernelization that outputs 
	a full kernel of
	$O(k^2)$ vertices and edges, or outputs that $(G, k)$ is a no-instance.
	The result can be used
	to find a (possibly not full) kernel 
	of at most $3k + 4$ vertices 
	in $O(n \log k + \poly(k))$ time using $O(\poly(k) \log n)$ bits.
\end{theorem}
\begin{proof}
In this proof we apply a modified BFS on the given graph, 
which adjustments are described below. 
The main structural adjustment is that the BFS
maintains at most $k+2$ quadruples instead of vertices in its queue
and uses the queue of the previous round to identify already visited vertices, instead
of marking all vertices of either visited or unvisited.
Recall that each quadruple $(v, p, i, v^*)$ consists of
a vertex $v$ and its predecessor $p$ if $v$ is not the root,
a counter $i \in \{0, \ldots, n\}$ with $i > 0$ being $v$'s position on a degree-2 chain,
and the vertex $v^*$ with $v^* \ne \op{null}$ being the $(k + 1)$th 
vertex on a degree-2 chain that contains~$v$.
During the run of the BFS we select vertices and edges that we can iteratively put
into a kernel under construction $G'$.

Before we start to describe the adjustments, we want to point out
that our approach works only if the BFS is started at a 
vertex of degree other than~$2$,
which we can identify by simply iterating over all vertices.
If there is no such vertex, then the graph is a simple cycle
and we output $G$ as the kernel if $m \le k + 2$, otherwise
we output ``no-instance''.

The BFS visits the vertices as usual and updates its quadruples as follows.
For each quadruple $(v, p, i, v^*)$, it iterates over~$v$'s neighborhood 
and stores the quadruple 
$(v', v, 1, \op{null})$ in queue $Q'$ for every unvisited neighbor~$v'$  if $v$ is of degree other than two, and otherwise
the quadruple $(v', v, i + 1, v^{**})$ where $v^{**}$ is $v$ if $i = k + 1$, otherwise $v^{**} = v^{*}$.

By Rule 2 we can bound the size of the BFS queue by $k+2$ and the size of the
kernel by $n' \le k^2 + 4k + 1 = O(k^2)$ vertices
and $m' \le n' - 1 + \frac{1}{2}(k^2 + 5k) = O(k^2)$ edges. To ensure Rule 2, we can
easily count the number of leaves in the BFS tree while executing the~BFS.

We now describe how a kernel $(G', k')$ can be constructed in adherence to Rule 1.
Instead of contracting arbitrary edges we
contract
edges
at the end of a degree-2 chain. The contraction is realized by not
copying
the inner vertices and edges at the end of a degree-2 chain
while the BFS traverses the paths 
and connecting the $(k+1)$st vertex with the last vertex of the path in the
kernel.
To avoid adding vertices into the queue that are 
already visited we
maintain the vertices of the previous and the current queue inside
two balanced heaps, respectively.

For the time being ignore a problem that 
two vertices in the BFS queue may be adjacent (i.e.,
the BFS
starts to explore a degree-2 chain from both its endpoints).
For each quadruple $(v, p, i, v^*)$, we add the vertex $v$ 
into $G'$ if $i \le k + 1$ and if additionally $p \ne \op{null}$, we also 
add the edge $\{v, p\}$ into $G'$---the condition ensures that we do not add
the full degree-2 chain into the kernel.
If the degree of $v$ in $G$ is not two,
then $v$ {\em terminates a degree-2 chain} and we 
add the edge $\{v^*, v\}$ if $i > k + 1$ (the bold edges in Fig.~\ref{fig:pc}). 
We additionally
add for every $u \in N(v)$ with~$u$ is in~$G'$, the
edge $\{v, u\}$ into $G'$ (in
Fig.~\ref{fig:pc} they are shown dashed).

We now consider the case where 
two vertices $v$ with $(v, p, i, v^*)$ and $v'$ with tuple $(v', p', i', v'^*)$ on the current
BFS queue are connected to each other in~$G$ and are both of degree two.
If $i + i' > k + 2$ we move backwards on both paths until $i + i' = k + 2$
(but $i, i' \ge 0$) and
modify the kernel by removing the vertices and edges used to move backwards.
Add the edge $\{w, w'\}$ to the kernel
where $w$ and $w'$ are the vertices 
at which we stopped our backward move. 

It remains to show the space and time bounds of our kernelization.
The size of the queues 
used for the BFS and our computation is bounded by $O(k)$ vertices and, thus, cannot exceed $O(k \log n)$ bits.
The kernel is bounded by $O(k^2)$ vertices and $O(k^2)$ edges and thus uses $O(k^2 \log n)$ bits.
In total, we use $O(k^2 \log n)$ bits.
Concerning the time bound note that a standard BFS runs in $O(n + m)$ time.
By Rule~2
$m = O(n + k^2)$ or we stop.
The algorithm has to check for each vertex if it is in a balanced heap (in the queue or in the kernel) of size at most $O(k^2)$, which takes $O(\log k)$ time per vertex.
In total we have running time of $O((n + k^2) \log k)$.
(Note that running backwards on degree-2 chains takes time linear to the length of the path
and thus our asymptotic time bound remains the same.)

Our kernel is small enough
to apply the polynomial-time kernelization of Li et al. and
we obtain so a kernel of $3k + 4$ vertices in $O(n \log n + \poly(k))$ time 
using $\poly(k) \log n$ bits. 
\end{proof}

\section{Log-Space Tree Traversal and Cycle Check}\label{sec:treeTraversalSection}
For our result on \textsc{Feedback Vertex Set} we require the following two auxiliary lemmas 
to traverse trees and find a back edge in a graph.
Cook and McKenzie~\cite{CookM87} showed how this can be done
in $O(n^2)$ and $O(n^3)$ time, respectively, by using $O(\log n)$ bits.

\begin{lemma}{(\cite[Theorem~2]{CookM87})}\label{lem:traverseTree}
	Given an $n$-vertex tree $T$ and a node $r$ of $T$ as root
	there is a $O(n^2)$-time $O(\log n)$-bits algorithm that
	traverses all vertices of $T$ in depth-first-search manner.
\end{lemma}
\begin{proof}
	Let $r \in T$ be a root and $p$ be the previously visited vertex.
	We use a known technique to traverse the graph in a special order,
	which main idea is as follows:
	Assume we visit a vertex $v$ from a vertex $p$.
	The vertex~$v$ has several neighbors $v_1, v_2, \ldots, v_i = p, \ldots, v_{\deg(v)}$, from which one of it is~$p$. Find the index $i$ of $p$ with $p = A[v][i]$ by iterating throw $v$'s adjacency array and visit ($A[v][(i + 1) \mod \deg(v)]$).
	If we return to $v$, then from the vertex $p' = v_{(i + 1)}$ and proceed with the next child $v_{(i + 2) \mod \deg(v)}$ of $v$. With the modulo operation we so visit all children  and leave the vertex via the back edge $(v, p)$.
 
	In detail we distinguish between the root (i.e., detectable via the check $v = r$) and the remaining nodes and treat them as follows.
		Let visit($v, p$) be the procedure to visit all nodes in depth-first-search manner.
		
	\begin{description}
		\item[Treat root] If $v = r \land deg(v) = 0$, we output $v$ and know that the tree consists only of one node, hence, we return.
			Otherwise and if $v = r \land p = null$ we know that its the first visit of the root, we output~$v$ and call visit($A[v][0]$) to visit its first child.
			Otherwise, we have returned to the root after visiting the maximal subtree below its child~$p$. We find the index $i$ of $p$ with $p = A[v][i]$ by iterating throw $v$'s adjacency array.
			Check if $p$ was the last child, i.e., ($i = \deg(v)$ holds) and return, since the whole tree was traversed.
			Otherwise, we call visit($A[v][i + 1], v$) to visit the next child.
		\item[Treat non-root] Output $v$.
		We find the index $i$ of $p$ with $p = A[v][i]$ by iterating throw $v$'s adjacency array and call visit($A[v][(i + 1) \mod \deg(v)]$).
	\end{description}
	
	Note, that the algorithm is actually defined recursively, however, it uses a tail-recursion which are known to be translateable to loops that do not need a stack. Hence, $O(\log n)$ bits suffice to store the required information.
	Concerning the running-time note that all operations except the search for the next edge to follow require constant time. The search itself requires $O(\deg(v))$ time and has to be done for each visit of a vertex, i.e., $O(\deg(v))$ times.
		Hence, the total required time is $\sum_{v \in T} \deg(v)^2 \le m(m + 1) = O(n^2)$ \cite[Theorem 3.7]{KinkarCh2003}, where $m$ is number of edges of a graph, which in case of a tree is bound by the handshaking lemma to $O(n)$. 
\end{proof}

\begin{lemma}{(\cite[Theorem~2]{CookM87})}\label{lem:isTree}
	Given an $n$-vertex graph $G$ and a vertex $r$, 
	there is an $O(n^3)$-time $O(\log n)$-bits algorithm 
	that either traverses the connected component with $r$ if it is a tree,
	or otherwise, it returns a back edge of the DFS tree rooted at~$r$.
\end{lemma}
\begin{proof}
Let us consider the scenario where the DFS, 
as described in Lemma~\ref{lem:traverseTree}, 
is on a path from $r$ to $u$ and is about to 
follow an edge $\{u, v\}$.
To find out if $\{u, v\}$ ''closes'' a cycle or not, 
the DFS must know if
$v$ has not been previously discovered by the DFS
or of $v$ is the direct predecessor of $u$ on the $r$-$u$
path.
Since the DFS has no knowledge on this, 
this condition is verified by a second DFS
run until reaching $u$
where we have to check for each
discovered vertex $v'$ if $v \ne v'$ and if
not true, if $v$ is not the direct predecessor 
of $u$ in the second DFS run.
Otherwise, this would imply a cycle constructed by
the paths $r$-$u$ and $r$-$v$ (where the first is not a
subpath of the second) and the edge $\{u, v\}$, implying
that the connected component is not a tree and $\{u, v\}$
is a valid back edge.

Since the algorithm can stop whenever a first cycle is found, 
the algorithm considers
only $O(n)$ edges. Thus
the space and time bounds stated in the lemma hold. 
\end{proof}

\section{Feedback Vertex Set}\label{sec:fvs}
Given an $n$-vertex $m$-edge graph $G = (V, E)$ a set $F \subseteq V$
is called {\em feedback vertex set} if
the removal of the vertices of $F$ from $G$ 
turns $G$ into an acyclic graph (also called {\em forest)}. 
In the parameterized \textsc{Feedback Vertex Set} problem,
a tuple $(G,k)$ is given where $G$ is a graph, and $k$ is a parameter.
We are searching for a feedback vertex set $F$ of size at most~$k$ in~$G$.
In kernelization, it is common to identify vertices that 
must be in every minimal feedback vertex set of size at most $k$, 
remove them from the instance, and restart the kernelization. 
To avoid modifying the given instance, 
we simulate this by starting with an empty set $F$ and 
adding vertices to $F$ when we determine they must be in 
every solution of size at most $k$. 
Subsequently, we realize the graph $G[V \setminus F]$ 
by considering $G$ and disregarding vertices in $F$.

Iwata showed a kernelization for \textsc{Feedback Vertex Set} 
that produces a kernel consisting of at most $2k^2 + k$ vertices 
and $4k^2$ edges and runs in $O(k^4 m)$ time~\cite{Iwata17}.
He mentions that all other kernelizations
for \textsc{Feedback Vertex Set} 
exploit an exhaustive application of
 the three basic rules below and the so-called $v$-flower rule.
A $v$-{\em flower of order $d$} is a set of $d$ 
cycles pairwise intersecting exactly on vertex~$v$.

\begin{description}
    \item[Loop Rule.] Remove a vertex $v$ with a loop and reduce to $(G - v, k - 1)$ and $F := F \cup \{v\}$.
	\item[Leaf Rule.] Remove a vertex~$v$ with $\op{deg}(v) \le 1$.
	\item[Chain Rule.] Remove a vertex~$v$ that has only
	two incident edges $\{v, u\}$ and $\{v, w\}$
	(possibly $u = w$), and add the edge $\{u, w\}$.
	\item[Flower Rule.] Remove a vertex $v$ if a $v$-flower 
	of order $k + 1$ exists and reduce to $(G - v, k - 1)$ and $F := F \cup \{v\}$.
\end{description}

By allowing $O(k \log n)$ bits for the algorithm, 
Elberfeld et al.\ also showed how to find a cycle of $2k$
vertices.
To realize the flower rule at a vertex $v$ we need to run along
up to $k + 1$ cycles and check if they
intersect at vertices other than~$v$.
If the given graph is reduced with respect to the Leaf and Chain Rule
and does not contain vertices with self-loops,
then it can be guaranteed that the smallest
cycle is of length at most $2k$ (maximum girth of a graph with the mentioned restrictions, minimum degree~3 and a feedback vertex set of size at most~$k$~\cite{DowneyF99}).
However, the length of the remaining cycles can be bound only by a function depending on $n$, not on $k$.
So it seems to be hard to find and verify a flower with $O(\poly(k) \log n)$ bits.
As shown by Iwata, one does not need the Flower Rule to find 
a kernel for FVS. Instead he uses 
a so-called
{\em $s$-cycle cover reduction}~\cite[Section 3]{Iwata17}
where he has to know
which edges incident to a vertex $s$ are bridges in the graph.
(For space bounds to find bridges, recall Section~\ref{sec:pc}.)
Since we have no solution to find a $v$-flower or an $s$-cycle with $O(\poly(k) \log n)$
bits, we show how to construct a kernel without using both rules exhaustively.

Thomass\'e~\cite{Tho10} introduced the rule below to compute a
kernel consisting of $4k^2$ vertices. As input we assume a simple graph.
Since his rule introduces
double edges,
our kernel $G'$ is a multi graph where every multi edge is a double edge.

\begin{description}
	\item[Thomass\'e's Rule~\cite{Tho10}] Let $X$ be a set of vertices, let
	$x \in V \setminus X$ and let $\mathcal C$ be a set of connected components  of $G \setminus (X \cup \{x\})$
	(not necessarily all the connected components) such that
	\begin{itemize}
		\item $G$ is loopless, with degree~$\ge 3$ and all multi-edges are double-edges,
		\item there is exactly one edge between $x$ and every $C \in \mathcal C$,
		\item every $C \in \mathcal C$ induces a tree, and
		\item for every subset $Z \subseteq X$, the number of trees of $\mathcal C$ having some neighbor in $Z$ is at least $2|Z|$.
	\end{itemize}
	Then reduce to $(G', k)$ where $G'$ is the graph obtained
	by joining $x$ to every vertex of $X$ by double edges and by removing the edges between $x$ and the components of $\mathcal C$.
\end{description}

Thomass\'e applies the rule to the whole graph~$G$. 
This means, he has to store graph changes over the whole graph~$G$. 
This is too expensive for us.
As discussed in the introduction, we utilize a separator $U$ 
to break down the graph into manageable components. 
This allows us to construct a kernel by processing and gradually 
incorporating these components. In the next subsection, 
we outline the construction of an approximate minimum feedback 
vertex set as separator $U$, so that the graph divides into 
trees.
In the subsequent subsection, we show how to iterate over the trees in $\mathcal T = G[V \setminus U]$. 
In a third subsection, we iteratively add these trees into a graph $G'$ (initially $G' = G[U]$) while upholding
our desired space bound of $G'$ having $O(k^4)$ vertices and edges, i.e., $O(k^4 \log n)$ bits.
However, the size of the trees in $\mathcal T$ is unbounded in $k$ 
and we need to perform an on-the-fly tree size reduction of the tree.
For this, we first make sure that every tree $T$ has not too many edges to $U$
(or we either find a vertex for the solution $F$ and restart, or conclude a no-instance).
Afterwards, we have to traverse $T$ and put
exactly those vertices of $T$ into $G'$ that are not removed by an
exhaustive application of the Leaf and Chain Rule.
To keep the size of $G'$ within our space bound
we show in a fourth subsection how to shrink $G'$ again.
To shrink the size of $G'$ to $O(k^2)$ vertices, we apply Thomass\'e's Rule, 
which has a precondition requiring that $G'$ has a minimum degree of~$3$. 
However, we cannot satisfy this precondition for vertices in $U$ within $G'$. 
Nevertheless, we can demonstrate that violating the precondition only 
for the vertices in $U$ still allows the rule to function if we 
adjust the bounds accordingly (see Lemma~\ref{lem:tree-reduction}).
Finally, we show that our construction of a kernel of $O(k^2)$ vertices 
and $O(k^3)$ edges runs in $O(n^5 \poly(k))$-time and with 
$O(k^4 \log n)$ bits with this approach. 

\smallskip
\noindent {\bfseries Separator $U$ of size $3k^2$.}
Becker and Geiger~\cite{BeckerG94} presented a $2$-approximation algorithm
for feedback vertex set in which they extend an $(2 \log d)$-approximation algorithm
(where $d$ is the maximum degree of the graph) by a phase that iteratively removes 
a vertex $v$ from the computed feedback vertex set~$S$ 
if all cycles that intersect~$v$ in $G$ also intersect $S \setminus \{v\}$.
It is unlikely that this can be done with $O(\poly(k)\log n)$ bits
or even with $O(f(k) \log n)$ bits
for some function $f$
since a cycle in $G$ can consists of $\Theta(n)$ vertices and there can be $\Theta(n)$ cycles.

We instead present only an $O(k)$-approximation algorithm, but it
runs with $O(\poly(k)\log n)$ bits. 
For this we use the 
following well-known rule.
Given a loopless graph $G$ of minimum degree~3 and without self-loops,
every FVS in $G$ of size at most $k$ contains at least one vertex
of the $3k$ vertices of largest degree~\cite[Lemma 3.3]{CyganFKLMPPS15}.
A graph with such properties can be computed by an exhaustive application of the Loop, Leaf and Chain Rule.
Elberfeld et al.~\cite[Theorem~4.13]{ElbST15}
showed how to implement the rules 
with $O(\log n)$ bits.
The graph obtained does not actually 
have to be stored. 
Instead we compute the required information
on demand with Lemma~\ref{lem:spaceBasicRule1and2}, 
which is similar to parts
of the proof of~\cite[Theorem~4.13]{ElbST15}.

\begin{lemma}\label{lem:spaceBasicRule1and2}
	Assume that an $n$-vertex $m$-edge graph $G = (V, E)$ and 
        a set $U \subseteq V$ consisting of $k^{O(1)}$ vertices is given.
        Let $G'$ be the graph obtained by an exhaustive application of 
        the Loop, Leaf and Chain Rule on $G[V \setminus U]$.  
	We can provide a structure that allows the iteration over the edges of every vertex of $G'$ 
	in $O(n^5 k \log k)$ total time
	by using $O(\log n)$~bits. In particular, we do not store $G'$.
\end{lemma}
\begin{proof}
Take $G_U = G[V \setminus U]$.
First of all, note that we can access $G_U$
(e.g., run a DFS in $G_U$) as if $G_U$ is given explicitly by accessing $G$
and ``ignoring'' all vertices (edges leading to vertices) in $U$.
More precisely, we define the neighbors $N_{G_U}(v) = N_G(v) \setminus U$.
Note that a Loop can be easily identified and possibly we restart.
Note further that the connected subgraphs that are removed from $G_U$ by an exhaustive application of 
the Leaf Rule are trees,
which we call {\em tree appendages}.
We can identify each neighbor $u \notin U$ of a vertex $v$
that is part of a tree appendage by 
running the algorithm of Lemma~\ref{lem:isTree}
on $G_U - v$ with $r = u$
as input.
If the algorithm returns that the connected component with $r$ in $G_U - v$ is a tree 
and does not visit~$v$, $u$ is part of a tree appendage
(not visiting~$v$ is important since otherwise, with~$v$ 
and the edge $\{v, u\}$ we have a cycle).
Let $Q_v$ be the set of neighbors of $v$ that are not part
of a tree appendage.
Let $G_1$ be the graph obtained from $G_U$
after an exhaustive application of the Leaf Rule.
Then $\deg_{G_1}(v) = |Q_v|$ is the degree of $v$ in $G_1$,
for every $v$ with $|Q_v| > 1$.
If $|Q_v| \le 1$, 
then $v$ is itself part of a tree appendage.
Otherwise,~$v$ is part of~$G_1$
and we can output all vertices of~$Q_v$
 as neighbors of~$v$
as required from the lemma.

Let $G_2$ be the graph obtained from $G_1$
after an exhaustive application of the Chain Rule.
Observe that, if $\deg_{G_1}(v) = 2$, 
then $v$ is part of a degree-2 chain in $G_1$ that is
replaced by an edge in $G_2$ by the exhaustive application of the Chain Rule.
A possibility is that the degree-2 chain
connects two vertices $u$ and $w$ of $G_1$
(possibly $u = w$) that are not of degree two.
Then $v$ is not part of $G_2$.
However, 
we cannot simple assume that every vertex $v$
with $\deg_{G_1}(v) = 2$
is not part of $G_2$ since there is 
a special case ($\star$):
$v$ may be
part of simple cycle consisting of only degree-2
vertices and the Chain Rule may reduce the cycle
to exactly one vertex $z$ with a self-loop.
The vertex $z$ can be an arbitrary vertex of
the cycle thus $z = v$ is possible.
We choose $z$ always as the
vertex with the smallest \textsc{id} of the cycle and ``ignore''
the remaining vertices. More precisely, if $v$ is such a
vertex, then $v$ is part of $G_2$
and has a self-loop.
If $\deg_{G_1}(v) > 2$, then~$v$ is part of~$G_2$,
but some of its edges in~$G_1$ may 
connect~$v$ with a 
degree-2 chain in~$G_1$ that is replaced by an 
edge in~$G_2$.

To realize the lemma iterate over each vertex $v$ of $G_U$:
If $Q_v \le 1$, ``ignore'' $v$.
If $\deg_{G_1}(v) = 2$, we output $v$ only if we are in the Special Case ($\star$)
and $v$ has the smallest \textsc{id} on its cycle.
We now may assume that 
$\deg_{G_1}(v) > 2$ and thus $v$ is part of $G_2$.
For each neighbor~$u$ of $v$ in $G_1$ follow
the potentially empty degree-2 chain from $u$ 
until a vertex $w$ with $w = v$ or 
$w \ne v \land \deg_{G_1}(w) > 2$.
If $w = v$, $v$ has a self-loop thus, 
we output~$v$ as a neighbor of~$v$.
If $w \ne v \land \deg_{G_1}(w) > 2$, 
we output $w$ as a neighbor of $v$.

Checking if $v$ is part of $G_1$ and outputting its neighbors can be 
done in $O(\deg(v)n^3 \log k)$ time: 
Lemma~\ref{lem:isTree} runs in $O(n^3)$ time per edge incident to~$v$.
However, since we have to ignore vertices and edges whose endpoints are in $U$ 
an access to $G_U$ needs an access to an heap of size $|U|$ and thus
runs in $\log k^{O(1)} = O(\log k)$ time.
We must iterate over all~$n$ vertices.
Thus, 
the total 
running time is 
$O(m n^3 \log k) = O(n^4 k \log k)$ (since $m = O(nk)$).

To check for the Chain Rule we have to additionally follow
degree-2 chains.
Since a chain can consists of $\Theta(n)$
vertices, this increases the running time by a factor of $n$,
resulting in a total running time of $O(n^5 k \log k)$ for the Chain Rule. 
\end{proof}

Since by Lemma~\ref{lem:spaceBasicRule1and2} we have access to a loopless graph of minimum degree three,
we iterate $k$-times over a graph $G[V \setminus U]$ (where initially $U = \emptyset$) 
and in each iteration select the $3k$ vertices of largest degree into $U$.
We so get an $O(k)$ approximate minimum feedback vertex set.

\begin{theorem}\label{thm:approxfvs}
	Given an $n$-vertex $m$-edge instance $(G, k)$ of \textsc{Feedback Vertex Set}, there is 
	an $O(n^5 k^2 \log k)$-time, $O(k^2 \log n)$-bits algorithm that
	either returns a feedback vertex set $U$ consisting of
	at most $3k^2$ vertices
	or answers that $(G, k)$ is a no-instance.
\end{theorem}
\begin{proof}
Starting with~$(G, k)$ and initially~$U = \emptyset$ 
as input we compute a graph~$G'$ by exhaustively
applying the Leaf and Chain Rule on~$G[V \setminus U]$.
If $G'$ is empty, we return~$U$ as a feedback vertex set for~$G$.
The Chain Rule may create self-loops and multi-edges.
If a vertex with a self-loop exists, it must be part of the minimal feedback vertex set, thus, put it into~$F$,
reduce~$k$ by one, and restart.
After an exhaustive application of the Leaf and Chain Rule 
the set consisting of $3k$ vertices of largest degree
contains at least one vertex of the minimal feedback vertex set~\cite[Lemma~3.3]{CyganFKLMPPS15}.
Thus, take $3k$ vertices of largest degree of $G'$ into $U$,
reduce $k$ by one, and restart.
If at any point $k < 0$, output ``no-instance''.
We so can compute a feedback vertex set~$U$ consisting
of at most $3k^2$ vertices.
Instead of storing $G'$ we compute the required information
with Lemma~\ref{lem:spaceBasicRule1and2}. 
We so iterate over each vertex $v$ of $G'$
and its edges
to determine its degree and check for self-loops
and compute~$U$ in $k$ rounds in
$k \cdot O(n^5 k \log k) = O(n^5 k^2 \log k)$ time
and with $O(|U| \log n) = O(k^2 \log n)$ bits. 
\end{proof}

\noindent {\bfseries Iterations over Trees.}
We want to output every tree $T$ of $G[V \setminus U]$ once.
For this, we iterate over all $u \in U$ 
and, intuitively speaking, output those trees~$T$ 
adjacent to $u$, i.e., every $T$ having a vertex $v$ such that 
$u$ and $v$ are adjacent in $G$. 
Note that with such an iteration we will not iterate over
components of $G$ that have no edges to any vertex of $U$. 
However, since those components are trees in $G$, and thus cycle free, we can ignore them.
Moreover, note that several vertices of $U$ can have edges to the same tree. 
We show in the proof of the next lemma how to avoid outputting a tree multiple times.  
To distinguish the trees, we
partition
the trees $\mathcal T$  
as follows (also see Fig.~\ref{fig:kindOfTrees}).
$\mathcal{T}_0$ is the set of trees in $\cal T$ that have at most one edge to a single vertex of~$U$.
$\mathcal{T}_1$ is the set of trees in $\cal T$ where each tree has at most one edge to at least two vertices of $U$.
$\mathcal{T}_2$ is the set of the remaining trees in $\cal T$
with least two edges to some vertex of~$U$.

\begin{figure}[h!]
\centering
	\begin{subfigure}{.3\textwidth}
		\centering
\begin{tikzpicture}[scale=0.9,transform shape,
	vertex/.style={draw,circle, fill=white, inner sep=0pt,
	minimum size=5pt}]
	
	\node[minimum size=0] (s) at (0, 0) {};

    \begin{scope}[]
		\node[vertex] (t1a) at ($(s) + (-40pt,50pt)$) {}; 
    	\node[vertex, fill=Plum] (t2a) at ($(t1a) + (-10pt,-10pt)$) {};
    	\node[vertex] (t3a) at ($(t1a) + (10pt,-10pt)$) {};
    	\node[vertex] (t4a) at ($(t2a) + (-10pt,-10pt)$) {};
    	\node[vertex] (t5a) at ($(t2a) + (10pt,-10pt)$) {};
		\draw (t1a) -- (t2a);
		\draw (t1a) -- (t3a);
		\draw (t2a) -- (t4a);
		\draw (t2a) -- (t5a);
	\end{scope}
	
    \begin{scope}[]
		\node[vertex] (t1b) at ($(s) + (0pt,50pt)$) {}; 
    	\node[vertex] (t2b) at ($(t1b) + (10pt,-10pt)$) {};
    	\node[vertex, fill=ForestGreen] (t4b) at ($(t2b) + (-10pt,-10pt)$) {};
    	\node[vertex] (t5b) at ($(t2b) + (10pt,-10pt)$) {};
		\draw (t1b) -- (t2b);
		\draw (t2b) -- (t4b);
		\draw (t2b) -- (t5b);
	\end{scope}
	
	\node[color=red, below = 10pt of t4b] (x) {$u$};
	\draw[dashed, color=red] (x) -- (t4b);
\end{tikzpicture}
		\caption{Trees in $\mathcal{T}_0$}	
	\end{subfigure}
	\begin{subfigure}{.3\textwidth}
		\centering
\begin{tikzpicture}[scale=0.9,transform shape,
	vertex/.style={draw,circle, fill=white, inner sep=0pt,
	minimum size=5pt}]
	
	\node[minimum size=0] (s) at (0, 0) {};
	
    \begin{scope}[] 
    	\node[vertex,fill=SkyBlue] (t2a) at ($(s) + (-10pt,55pt)$) {};
    	\node[vertex] (t3a) at ($(t2a) + (20pt,0pt)$) {};
    	\node[vertex] (t4a) at ($(t2a) + (10pt,-6pt)$) {};
		\draw (t2a) -- (t3a);
		\draw (t2a) -- (t4a);
	\end{scope}
	
    \begin{scope}[]
		\node[vertex,,fill=ForestGreen] (t1b) at ($(s) + (-10pt,40pt)$) {}; 
    	\node[vertex] (t2b) at ($(t1b) + (10pt,-5pt)$) {};
    	\node[vertex] (t4b) at ($(t2b) + (-10pt,-10pt)$) {};
    	\node[vertex] (t5b) at ($(t2b) + (10pt,-10pt)$) {};
		\draw (t1b) -- (t2b);
		\draw (t2b) -- (t4b);
		\draw (t2b) -- (t5b);
	\end{scope}
	
    \begin{scope}[]
		\node[vertex,fill=Plum] (t1c) at ($(s) + (0pt,10pt)$) {}; 
	\end{scope}
	
	\node[color=red] (x1) at ($(s) + (-30pt,50pt)$) {$u_1$};
	\node[color=red] (x2) at ($(s) + (30pt,50pt)$) {$u_2$};
	\node[color=red] (x3) at ($(s) + (-30pt,25pt)$) {$u_3$};
	\node[color=red] (x4) at ($(s) + (30pt,25pt)$) {$u_4$};
	\node[color=red] (x5) at ($(s) + (-30pt,10pt)$) {$u_5$};
	\node[color=red] (x6) at ($(s) + (30pt,10pt)$) {$u_6$};
	
	\draw[dashed, color=red] (x1) -- (t2a);
	\draw[dashed, color=red] (x1) -- (t1b);
	\draw[dashed, color=red] (x3) -- (t2b);
	\draw[dashed, color=red] (x3) -- (t1c);
	\draw[dashed, color=red] (x5) -- (t1c);

	\draw[dashed, color=SkyBlue] (x2) -- (t3a);
	\draw[dashed, color=SkyBlue] (x2) -- (t1b);
	\draw[dashed, color=ForestGreen] (x4) -- (t5b);
	\draw[dashed, color=ForestGreen] (x4) -- (t1c);
	\draw[dashed, color=Plum] (x6) -- (t1c);
\end{tikzpicture}
		\caption{Trees in $\mathcal{T}_1$}
	\end{subfigure}
	\begin{subfigure}{.3\textwidth}
		\centering
\begin{tikzpicture}[scale=0.9,transform shape,
	vertex/.style={draw,circle, fill=white, inner sep=0pt,
	minimum size=5pt}]

	\node[minimum size=0] (s) at (0, 0) {};

    \begin{scope}[]
		\node[vertex] (t1a) at ($(s) + (-45pt,30pt)$) {}; 
    	\node[vertex] (t2a) at ($(t1a) + (-10pt,-5pt)$) {};
    	\node[vertex] (t3a) at ($(t1a) + (10pt,-5pt)$) {};
    	\node[vertex,fill=YellowOrange] (t4a) at ($(t2a) + (0pt,-10pt)$) {};
    	\node[vertex] (t5a) at ($(t2a) + (10pt,-5pt)$) {};
		\draw (t1a) -- (t2a);
		\draw (t1a) -- (t3a);
		\draw (t2a) -- (t4a);
		\draw (t2a) -- (t5a);
	\end{scope}

    \begin{scope}[]
		\node[vertex] (t1b) at ($(s) + (0pt,30pt)$) {}; 
    	\node[vertex] (t2b) at ($(t1b) + (10pt,-5pt)$) {};
    	\node[vertex] (t4b) at ($(t2b) + (-10pt,-5pt)$) {};
    	\node[vertex,fill=ForestGreen] (t5b) at ($(t2b) + (0pt,-10pt)$) {};
		\draw (t1b) -- (t2b);
		\draw (t2b) -- (t4b);
		\draw (t2b) -- (t5b);
	\end{scope}

	\begin{scope}[]
		\node[vertex] (t1c) at ($(s) + (-20pt,30pt)$) {}; 
    	\node[vertex,fill=Plum] (t2c) at ($(t1c) + (10pt,0pt)$) {};
		\draw (t1c) -- (t2c);
	\end{scope}

    \begin{scope}[]
		\node[vertex] (t1d) at ($(s) + (-40pt,0pt)$) {}; 
    	\node[vertex] (t2d) at ($(t1d) + (10pt,-10pt)$) {};
    	\node[vertex,fill=SkyBlue] (t3d) at ($(t1d) + (-10pt,-10pt)$) {};
		\draw (t1d) -- (t2d);
		\draw (t1d) -- (t3d);
	\end{scope}

    \begin{scope}[]
		\node[vertex] (t1e) at ($(s) + (0pt,0pt)$) {}; 
    	\node[vertex] (t2e) at ($(t1e) + (0pt,-10pt)$) {};
    	\node[vertex,fill=red] (t4e) at ($(t2e) + (-10pt,0pt)$) {};
		\draw (t1e) -- (t2e);
		\draw (t2e) -- (t4e);
	\end{scope}

	\node[color=red] (x) at ($(s) + (-20pt,10pt)$) {$u$};
	
	\draw[dashed, color=red] (x) -- (t3a);
	\draw[dashed, color=red] (x) -- (t5a);
	
	\draw[dashed, color=red] (x) -- (t1c);
	\draw[dashed, color=red] (x) -- (t2c);
	
	\draw[dashed, color=red] (x) -- (t4b);
	\draw[dashed, color=red] (x) -- (t5b);
	
	\draw[dashed, color=red] (x) -- (t4b);
	\draw[dashed, color=red] (x) -- (t5b);
	
	\draw[dashed, color=red] (x) -- (t1d);
	\draw[dashed, color=red] (x) -- (t2d);
	
	\draw[dashed, color=red] (x) -- (t1e);
	\draw[dashed, color=red] (x) -- (t4e);
\end{tikzpicture}
		\vspace{10pt}%
		\caption{Trees in $\mathcal{T}_2$}
	\end{subfigure}
\caption{Our partition of trees with edges to $U=\{u_1,u_2,u_3,\ldots\}$. 
The colored vertices are the vertices
with the smallest \textsc{id} in the trees.}
\label{fig:kindOfTrees}
\end{figure}

\begin{lemma}\label{lem:presentTreesOnce}
	Given an $n$-vertex $m$-edge graph $G = (V, E)$ and a 
	set $U$ of $O(k^2)$ vertices, 
	there is an algorithm that outputs a single vertex $w$ of $T$ 
	as a representative for each tree $T$ in $\mathcal T_1 \cup \mathcal T_2$
	and some trees of $\mathcal T_0$
	once. The algorithm runs in $O(n^3 k^3 \log^2 k)$ time and with $O(\log n)$ bits.
\end{lemma}
\begin{proof}
	Let a vertex be smaller than another vertex, if its vertex label is smaller.
	For each vertex $u$ in $U$, iterate over its neighbors $w$ not in $U$. 
	For each such $w$, consider it as the root of a tree $T$. 
	Traverse through $T$ to find the smallest $u'$ in $U$ connected to $T$. 
	If $u'$ is not $u$, continue the iteration over $U$ with the next vertex
	since $T$ was previously 
	acknowledged as adjacent to a smallest vertex in $U$. 
	If not, output $w$ as a representative for $T$.

	We now focus on the performance of the algorithm.
	The iteration over $U$ takes $O(|U|)$ time.
	The iteration over the neighbors of $U$ ignoring neigbors in $U$ 
	increases the time
	by a factor of $O(n \log |U|)$ to $O(|U| n \log |U|)$.
	Possibly each vertex is connected to a ``large'' tree, whose
	traversal with an adjusted Lemma~\ref{lem:traverseTree} can be done
	in $O((n^2 \cdot nk) \log |U|)$ time
	increasing the time to $O(|U| n^3 k \log^2 |U|)$.
	(The adjustment is necessary to ignore edges to $U$, which requires an
	$O(\log |U|)$-time membership check in $U$. Furthermore, since we have to
	consider edges to $U$ the time increases in each step at most by the possible number of edges, which is bounded by
	$O(nk)$ for every yes-instance of FVS.)
	Note that the time to find the smallest vertex in $U$ is already
	included in the time since we already have to deal with vertices 
	in $U$ by performing membership checks.
	Summarized, the running time is $O(|U| n^3 k \log^2 |U|) = O(n^3 k^3 \log^2 k)$.
	Considering the space, we need a constant amount of
	local variables and apply Lemma~\ref{lem:traverseTree} 
	which sums up to $O(\log n)$ bits.  
	
\end{proof}

Observe that we do not need to add trees 
of $\mathcal{T}_0$ to $G'$ since they 
can be removed by the Leaf Rule anyway.
\mbox{If we identify such a tree, we skip over it.} \smallskip

\noindent {\bfseries Tree Size Reduction.}
We now want to shrink each tree $T$ of $G[V \setminus U]$ so that we can add them to
$G'$ without exceeding our space bound of $O(k^4 \log n)$ bits.
Due to Cook and McKenzie~\cite{CookM87}, $O(\log n)$ bits
DFS exists which suffice to find out for each tree $T$
to which set $\mathcal{T}_0$, $\mathcal{T}_1$ or $\mathcal{T}_2$ it belongs.

By definition a tree $T \in \mathcal T_1$ can have at most one edge
to every vertex of $U$. 
Thus $T$ has at most $|U| = O(k^2)$ edges into $U$.
By the lemma below, we can add $T$ into $G'$.

\begin{lemma}\label{lem:addTree}
	Given $U$ and an $\bar{n}$-vertex tree $T=(V_T,E_T)$ in $G[V\setminus U]$ 
	such that $T$ has $\ell$ edges to $U$.
    After applying the Leaf and Chain Rule to $G[V_T \cup U]$ while forbidding the removal of vertices of $U$, 
    $T$ has at most $O(\ell)$ vertices. This can be done in $O(\bar{n}^3)$ time using $O(\log n)$ bits.
\end{lemma}
\begin{proof}
	Traverse the tree $T$ with the algorithm of Lemma~\ref{lem:traverseTree}.
	Before visiting a new vertex~$v$ from a vertex $u$, check with Lemma~\ref{lem:traverseTree}
	if a vertex of the subtree $T'$ with root~$v$ is adjacent with some vertex of $U$.
	If not, skip over $v$ and thereby the whole subtree $T'$ (since $T'$
	can be removed by the Leaf Rule).
	Otherwise check if a chain starts at $v$ and ends at $w$. If so,
     	add only the edge $\{u,w\}$ to $G'$ and continue at
	vertex $w$ ($v$ is removed by the Chain Rule). If not, add vertex $v$ as well as edge $\{u,v\}$ and
	continue with the children of $v$. Let $T'$ be the final tree.

        Let $\bar{n}$ be the vertices of $T$. Then the traversal over all
        $\bar{n}$ subtrees $T$ runs in $\bar{n}\cdot O(\bar{n}^2)=O(\bar{n}^3)$ time.
        To bound the vertices $n'$ of $T'$, note that $T'$ has $n'-1$ edges (total degree is $2n'-2$), but every vertex of $T'$
        must have degree at least three if we add the edges to $U$. Thus,
        $3n'<=2n'-2+\ell$ and $n'<=\ell-2$. 
\end{proof}

For a tree $T = (V_T, E_T) \in \mathcal T_2$, 
a vertex of $U$ can be connected
to multiple vertices of $T$.
Hence we need to run the following two steps to bound the number of edges,
where Step~1 is a precondition for Step~2.
By Step~2 we get subtrees $T'$ of $T$ such that $T'$ and $U$ have at most $(k + 1)^2$ edges in between. Thus,
we can add $T'$ with $O(k^2)$ vertices to $G'$ by Lemma~\ref{lem:addTree}.

\begin{description}
	\item[Step 1:] Consider a bipartite graph $Y = (U \cup V_T, E')$ where $\{u, w\} \in E'$ exactly if there are at least $k + 2$ internally vertex disjoint paths
        between $u$ and $w$. (For the ongoing algorithm observe: To find a solution for feedback vertex set of
	size $k$ in $Y$, we need a vertex cover of size at most $k$ in $Y$.
	Furthermore, any vertex in any vertex cover of size $k$ in $Y$ 
	must also be in any feedback vertex set of size $k$ in $G$.)
	If a vertex $u \in U$ has degree at least $k + 1$ in $Y$, take $u$ into
        our solution set $F$ and restart.
        If $Y$ has more than $k^2+k$ vertices, 
        output no-instance.
        Otherwise, define the common vertices of $Y$ and $V_T$ as set
        $U'$. Note that $|U'| \le k^2$. Temporarily take $U\cup U'$ as separator. This splits $T$ in several 
        {\em small trees}~$T'$,
        which we will process iteratively in Step 2.
	
	\item[Step 2:] If a small tree $T'$ has at 
	least $(k + 1)^2$ edges to a vertex 
	$u \in U \cup U'$, take $u$ into our
	solution $F$ and restart. 
	Otherwise add $T'$ to our graph $G'$.
\end{description}

\begin{lemma}\label{lem:stepsSafe}
	Steps 1 and 2 are safe and both steps run in $O(n^3 \log k)$ time and with $O(k^3 \log n)$ bits.
\end{lemma}
\begin{proof}
Consider
Step 1. A vertex $u$ of degree $k+1$ in $Y$ must be in any vertex cover of size
$k$ and, by construction, the vertex must be also in our feedback vertex
set, i.e., it is safe to add $u$ to $F$. Furthermore, if $Y$ has more than
$k^2+k$
vertices, but degree bounded by $k$, then $Y$ has no vertex cover of size~$k$ and thus
we can not find a feedback vertex set of size~$k$. To sum up, Step 1 is safe.

Now focus on Step 2.
Let $X$ be a set of the vertices of $T$ that are adjacent to a fixed vertex $u \in U
\cup U'$. 
Ignoring the parts of $T$ that are not on a path between two vertices of $X$, 
we obtain a tree with maximum degree $\Delta \le k + 1$ by Step 1.
One can easily see that, 
given a tree $T = (V_T, E_T)$ with maximum degree $\Delta$ and $X \subseteq V_T$,
we can find $\lfloor |X|/(\Delta + 1)\rfloor$ pairs of
vertices in $X$ such that the paths in $T$ between each pair are
vertex disjoint~\cite[Lemma 2.4]{ErlKLSS19}.
Thus, if $u$ has $|X| \ge (k + 1)^2$ edges to $T$, we have an $x$-flower of order $|X|/(k + 1) \ge k + 1$.
Thus, also Step 2 is safe. 

We now turn our attention to the algorithmic details of Step 1.
To track the number of internally vertex-disjoint paths between vertices, 
we employ a table $C$. Each vertex of $U$ gets its own counter in this table.

As we traverse the tree $T$ by Lemma~\ref{lem:isTree}, consider every vertex
$w$ once in $T$. 
Every neighbor of $w$, denoted as $v_i = v_1, v_2, \ldots, \deg(w)$, 
serves as the roots of a maximal subtree $T - w$.
Observe here that if $\ell$ subtrees have edges leading to a vertex $u \in U$ in graph $G$, 
then $\ell$ internally vertex-disjoint paths exist between $w$ and $u$. 

Now, for each subtree rooted at $v_i$, traverse it. 
Whenever an edge leads to a vertex $u \in U$, set the corresponding value $C'(u)$ 
to true in a temporary table $C'$ (initialized with false). 
After completing the subtree traversal, increase the counter in 
$C$ for every vertex in $C'$ marked true.

If any vertex $u$ in $C$ sees its counter surpass $k + 1$, insert the edge 
$\{u, w\}$ into an initially empty graph~$Y$. 
Whenever a vertex gets degree greater than $k + 1$ in $Y$ put into $F$ and restart.
(The details are explained in Step 1). Moreover, if $Y$ exceeds $k^2 + k$ vertices,
output no-instance.
At the end, construct $U'$ as directed in Step~2.

Concerning the running time, an applications of Lemma~\ref{lem:isTree} 
in every tree 
runs in $O(n^3 \log |U|)$ total
time. Since we then iterate over each tree for each
vertex $w$ once, i.e., at most
$O(n)$ times, it can be done 
by Lemma~\ref{lem:traverseTree} in $O(n^2 \log |U|)$ total time where the
extra factor of $\log |U|$ comes from membership tests in $U$. In the same
time, we can compute $U'$. To sum up, Step 1 runs in $O(n^3 \log k)$
time.
Concerning our space consumption, we can easily observe that $Y$ has never
more than $O(k^2)$ vertices and $O(k^3)$ edges, i.e., $O(k^3 \log n)$ bits
suffice.

We finally consider the time and space necessary for Step 2. We can easily
count the number of edges of a tree $T'$ to $U$ by Lemma~\ref{lem:traverseTree}
in neglitable time and space. By Lemma~\ref{lem:addTree}, we can simply add $T'$ into
$G'$. 
\end{proof}

Finally note that, by Step 2 above, we temporarily have a separator $U\cup U'$ of size $4k^2$.
After adding all subtrees $T'$ of a tree $T$ in $\mathcal T_2$ to $G'$, we can also
add $U'$ to $G'$ and we are back to a separator of size $3k^2$.
\smallskip

\noindent{\bf{Shrink the Kernel again.}}
After adding several trees to $G'$
we have to ensure that the size of $G'$ does not exceed
our space bound. We run into two issues.
\begin{enumerate}
	\item A naive application of a reduction rule may ``unsafely'' remove 
vertices since only subgraphs are considered, e.g.,
a vertex of $U$ could be considered as being a leaf in $G'$ because
several vertices that are connected to $U$ were not added
to $G'$ yet.
	\item Thomass\'e's Rule requires $G'$ to be 
a loopless graph of minimum degree~3, and we
cannot ensure that the vertices of~$U$ in $G'$ are of minimum degree~3.
\end{enumerate}

We address issue (1) with respect to all reduction rules.
The application of the Loop and the Flower is still safe to use
because whenever they apply, vertices are selected into a solution $F$ for feedback vertex set
and we restart. To deal with the Leaf and the Chain Rule we forbid that vertices of $U$
are removed in $G'$. 
Thomass\'e's Rule does not remove vertices.
Instead, it only removes existing edges and adds new ones within our subgraph~$G'$.
Moreover as stated in the rule, it does not have to consider all connected components,
which makes the rule safe for usage in $G'$.

Concerning issue (2) we show in the next lemma that vertices of $U$ can be exempted from
fulfilling the property of being of minimum degree~3.

\begin{lemma}\label{lem:tree-reduction}
Let $G = (V, E)$ be is a subgraph of $G'$ where 
$G'$ is a loopless graph and multi-edges are double edges,
such that $G$ has $n > 16k^2$ vertices and
only a subset $U \subseteq V$ with $|U| \le 4k^2$
have degree 0, 1 or 2 in $G$. 
If $G$ has a feedback vertex set of size $k$, then we can apply the Flower Rule or Thomass\'e's Rule in polynomial time and with $O(k^2 \log n)$ bits.
\end{lemma}
\begin{proof}
     Let $S$ be a feedback vertex set of $G$ with at most $k$ vertices.
     Then $G[V \setminus S]$ is a forest and has at most $|V \setminus S|-1$ edges. 
     Take $U' := V \setminus U$. 
     Since all vertices in $U'$
     have degree at least 3, the total number of edges 
     between $S$ and $V \setminus S$ is at least
     $3|U' \setminus S| - 2|V \setminus S| -1 
     = 3 \cdot ((6k^2-4k^2)-k) - 2 \cdot (6k^2 - k) -1 =
     36k^2-3k - 32k^2+2k -1 > 4k^2 - k$.
     Thus, a vertex of $S$ has degree at least $4k$.

     Now Thomass\'e concludes~\cite[Theorem 4.1]{Tho10} that we
     find in polynomial time a flower of order $k+1$ or we can apply Thomass\'e's Rule to $G$. 
\end{proof}

\noindent {\bfseries Construct a kernel:}
By Theorem~\ref{thm:approxfvs}, we have access to a approximate minimum feedback 
vertex set $U$ of size $3k^2$. $U$ divides $G$ into several trees, enabling us
to construct our initial kernel $G' = G[U]$ under construction.
We iterate over each tree $T$ of $\mathcal T_1 \cup \mathcal T_2$ using Lemma~\ref{lem:presentTreesOnce} and determine its type.
If $T \in \mathcal T_1$, we add it to $G'$ using Lemma~\ref{lem:addTree}.
Otherwise, if $T \in \mathcal T_2$, it must be integrated into $G'$ by first breaking it
down into smaller trees. To achieve this, we execute Step 1 and 2 
and get another separator $U'$.
Combining $U'$ with $U$ devides $T$ into several subtrees $T_1, T_2, \ldots$.
We then utilize Lemma~\ref{lem:presentTreesOnce} with $G[V \setminus (U \cup U')]$, but use $U'$
to iterate over the trees $T_i = T_1, T_2, \ldots$ (since they are all adjacent to $U'$)
and add one tree at a time with Lemma~\ref{lem:addTree}.
While adding a tree apply the Loop, Leaf, Chain Rule and shrink $G'$ by the Flower and Thomass\'e's Rule.
If all trees of $T_i$ were added, add also $U'$ into $G'$.
After all trees of $G$ have been added to $G'$, we apply the best 
kernelization~\cite{Iwata17} to get a kernel of $2k^2 + k$ vertices.

\begin{theorem}\label{th:fbs}
	Given an $n$-vertex instance $(G, k)$ of \textsc{Feedback Vertex Set}, there is 
	an $O(n^5 \poly(k))$-time, $O(k^{4} \log n)$-bits
	kernelization that either outputs a kernel consisting
	of $2k^2 + k$ vertices or returns that $(G, k)$ is a no-instance.
\end{theorem}
\begin{proof}
	Since we shrink $G'$ (initial size $|U|=O(k^2)$)
         repeatedly to $O(k^2)$ vertices after adding trees
	consisting of $O(k^4)$ vertices and edges we have a space bound
	of $O(k^4 \log n)$ bits. All other algorithms also run within this
	space bound. 

	Concerning the running time,
	the construction of $U$ runs with Theorem~\ref{thm:approxfvs} in $O(n^5 k^2 \log k)$.
	Even if we restart at most $k$ times we can reuse $U$ by removing vertices from~$U$
	whenever they become part of~$F$.

	An iteration over all trees by Lemma~\ref{lem:presentTreesOnce} runs in $O(n^3k^3\log^2 k)$ time.
	To identify the type of a tree, we have to traverse it with Lemma~\ref{lem:traverseTree} which runs in
	$O(n^2 \log k)$ time (while performing membership checks for $U$ (and $U'$)).
	To perform Step 1 und 2 for each tree we have to run Lemma~\ref{lem:stepsSafe},
	which can be done in $O(n^3 \log k)$ time.
	Adding a tree to $G'$ runs with Lemma~\ref{lem:addTree} in $O(n^3)$ time.
	Applying the kernelization rules on $G'$ runs in $O(\poly(k))$ time.
	
	Since we have at most $n$ trees, the total running time is $O(n^5 k^2 \log k) +
	O(n^3k^3\log^2 k) + O(n) \cdot (O(n^2 \log k) + O(n^3 \log k) + O(n^3) + O(poly(k)))$,
	which in total is $O(n^5 \poly(k))$. 
\end{proof}

\section{Cluster Editing and Cluster Deletion}\label{sec:ce}
The \textsc{Cluster Editing} problem can be described as follows. 
Given a graph $G=(V,E)$ with $n$ vertices and $m$ edges, 
and an integer parameter $k$, can we add or delete 
no more than $k$ edges such that the modified graph comprises entirely of disjoint cliques?
Recall that due to Bannach et al.~\cite{BanST15} \textsc{Cluster Editing}
is in $\paraL$ (more precisely a subclass of it).
Our goal is to address the need of Heeger et al.~\cite{HegHKNRA21}
for a space-efficient full kernel
for \textsc{Cluster Editing}, which is used
in their framework
in a temporal setting of
the cluster editing problem.
To solve the problem, it is important to find so-called
 {\em conflict triples} in $G$. Each conflict tripple
 is a subgraph formed by vertices $\{u, v, w\}$ 
with edges $\{u, v\}$ and $\{v, w\}$, but lacking an edge $\{w, u\}$. 
For each existing conflict triple, one should either 
remove one of the edges $\{u, v\}$ or $\{v, w\}$ 
or add a new edge $\{w, u\}$.

In the folklore technique for \textsc{Cluster Editing} kernelization, 
we iterate over all conflict triples and, 
for every vertex pair ${u, v}$ within a conflict triple, 
we maintain two global counters: $C_{u, v}$ for the occurrence 
of the edge $\{u, v\}$ and $C'_{u, v}$ for the conflict triples 
where the edge $\{u, v\}$ is missing.
Following this iteration, we update the graph as follows: 
add an edge $\{u, v\}$ to the graph if it is missing in at least $k + 1$ 
conflict triples (i.e., for each pair ${u, v}$ with $C'{u, v} \ge k + 1$
add the edge $\{u, v\}$), 
and remove all edges $\{u, v\}$ that are part of at least $k + 1$ 
conflict triples (i.e., all edges $\{u, v\}$ with $C_{u, v} \ge k + 1$). 
We then reset all counters and repeat the described iteration process.

After each iteration, 
we count the number of non-zero counters in $C$ and $C'$. 
If this count exceeds $(k+1)^2$, we conclude with a "no-instance" answer 
(since adding or deleting a single edge $\{u, v\}$ can resolve at most $k + 1$ 
conflict triples in~$G$). Otherwise, non-zero counters indicate exactly 
those vertices that are not part of a clique. For a full kernel, we include at 
most $(k+1)^2$ conflict triples, encompassing both vertices and edges. 
If, across all iterations, an edge is introduced and then removed 
(or vice versa), we answer with ``no-instance''.

Typically, a counter is created for every conflict tripple. However, since
only non-zero counters are of interest, we maintain only those counters and
reject instances as ``no-instance'' as soon as they
exceed $(k+1)^2$ different counters.

\begin{theorem}
Given an $n$-vertex $m$-edge instance $(G, k)$ of \textsc{Cluster Editing},
there is an 
$O(nm\log k)$-time $O(k^2\log n)$-bits
kernelization that either outputs
a full kernel 
of $O(k^2)$ vertices/edges or returns that $(G, k)$ is a no-instance.
\end{theorem}
\begin{proof}
We next describe and analyze an implementation of the algorithm above. 
First of all, we can iterate over all vertex triples $\{u,v,w\}$
by iterating over all edges and, for each edge $\{u,v\}$, over all vertices~$w \in N(v) \cup N(u)$.
We cannot modify $G$ on the read-only word-RAM, instead
we store $O(k)$ modifications to $G$ in a heap and whenever there is an access to $G$, we
	access the heap to check for the existence of an edge, for which we pay for
with an extra factor in the running time logarithmic to the size of the heap.
Thus, we can iterate over the vertex triples of the virtually modified graph $G$
in $O(m \log k)$ time.
The time needed to update all $O(k^2)$ counters in a heap 
is $O(k^2 \log k)$---note that we can assume without loss of generality that $k<m$ or we have a yes-instance.
Since the iteration has to be repeated at most $(k + 1)$ times, the final running time
is $O(m k \log k)$.
The time $(k^2 \log k)$ to create the full kernel by adding $(k+1)^2$ vertex pairs and edges 
to an initially empty graph is included in the final running time.

We have to store the counters, but only the non-zero counters, the modifications to $G$
in a heap and the kernel.
Because the amount of non-zero counters is bounded by $O(k^2)$, the number of modifications
is bounded by $O(k)$, and the kernel contains $k(k + 2) = O(k^2)$ vertices
and $k(2k + 1) = O(k^2)$ edges
and therefore, 
our algorithms uses $O(k^2 \log n)$ bits.
The full-kernel consists of $k(k + 3)$
vertices and $k(2k + 3)$ edges. 
\end{proof}

Note that to solve \textsc{Cluster Deletion} the only change is that we 
return a no-instance for $(G,k)$ whenever a counter
$C'_{u,v}\ge k+1$ for an edge $\{u,v\}$.

\bibliography{main}

\end{document}